# Repeated magnetic compensation behaviour in $Nd_{0.75}Gd_{0.25}Rh_3B_2$ alloy having hexagonal structure and planar anisotropy

Prasanna D. Kulkarni*, U. V. Vaidya, V. C. Rakhecha, A. Thamizhavel, S. K. Dhar, A. K. Nigam, S. Ramakrishnan and A. K. Grover[†]

Department of Condensed Matter Physics and Materials Science,

Tata Institute of Fundamental Research, Colaba, Mumbai, India 400005.

[*]email: prasanna@tifr.res.in

[†]email: grover@tifr.res.in

## Abstract

The results of dc and ac magnetization, electrical resistance and heat capacity studies in an admixed rare earth alloy $Nd_{0.75}Gd_{0.25}Rh_3B_2$, which is close to the zero magnetization limit, are reported. Novel observations include the two antiferromagnetic like transitions, concomitant repeated/multiple magnetic compensation behaviour and a characteristic oscillatory magnetic response, with temperature dependence of magnetization at low fields crossing the M = 0 axis thrice. The effective coercive field relating to the residual remanence also has an oscillatory response, with two minima located close to two 'zero-crossover' temperatures. The unusual behavior in $Nd_{0.75}Gd_{0.25}Rh_3B_2$ is conjectured to arise from its anisotropic hexagonal structure, with planar anisotropy, resulting in a *quasi* one dimensional electronic structure and the accentuated role of conduction electron polarization near the zero magnetization limit.

PACS numbers: 75.50.Ee, 71.20.Lp, 75.47.-m, 75.40.-s

**I. Introduction**

Rare-earth (R) elements of the 4f-series in the periodic table easily combine with nonmagnetic elements[1] to form different series of isoelectronic intermetallic compounds, which usually undergo ferromagnetic or antiferromagnetic ordering, depending on the conduction electron mediated indirect exchange interaction between the spins of the $R^{3+}$ ions. In a ferromagnetic series, one can conceive admixed R-alloys[2], with a given R site randomly occupied by two $R^{3+}$ ions ($R_1$ and $R_2$) belonging to different halves of the 4f-series, for which the total angular momentum J in the ground state of the spin-orbit (S-L) multiplet structure has eigenvalues, J = L-S and L+S, respectively. In an admixed alloy belonging to a ferromagnetic series, the magnetic moments of $R_1$ and $R_2$ ions ($\mu_{R1}$ and $\mu_{R2}$) are observed to be antiferromagnetically coupled[2,3,4] and at specific stoichiometries (e.g., when (1-x) $\mu_{R1}$ ~ (x) $\mu_{R2}$)), one usually comes across field induced moment reversal phenomenon[2,3,5-7], the basis of which is the occurrence of a magnetic compensation (at low fields)[5-8] between the opposing contributions from $\mu_{R1}$ and $\mu_{R2}$. Bulk magnetic response in a metallic sample also has a contribution from spins of conduction electrons, which remains in phase with those from the respective spin parts of $R^{3+}$ ions. Thus, for admixed rare earth alloys relating to a given ferromagnetic series, the notion of spin ferromagnetism prevails all through. The conduction electron polarization (CEP) contribution, though small, can assume accentuated importance[9] in admixed stoichiometries very close to the limit, (1-x) $\mu_{R1}$ ≈ (x) $\mu_{R2}$[5,8,10-13], where the samples are expected to yield near zero magnetization response.

We report here novel observations of (i) repeated/multiple compensation behaviour in the magnetization response and (ii) an oscillatory feature in the coercive field versus temperature plot in a spin ferromagnet, $Nd_{0.75}Gd_{0.25}Rh_3B_2$. This stoichiometry, estimated to have nominal zero magnetization, is observed to display crossing of the M = 0 axis thrice at low fields. The stoichiometry $Nd_{0.75}Gd_{0.25}Rh_3B_2$, made up of two ferromagnetic compounds, $NdRh_3B_2$ ($T_c$ ~ 10-15 K, $\mu/Nd^{3+}$ ~ 2.5 $\mu_B$) and $GdRh_3B_2$ ($T_c$ ~ 93 K, $\mu$/f.u. ~ 7.7 $\mu_B$)[14,15], crystallizes in the hexagonal $CeCo_3B_2$ type structure, in which the Nd/Gd ions occupy a unique crystallographic site and are stacked in hexagonal layers[15]. We recall that for the free $Nd^{3+}$ ions, S = 3/2, L = 6, J = 9/2 and $g_J$ = 8/11, and for $Gd^{3+}$ ions, S = 7/2, L = 0, J = 7/2 and $g_J$ = 2. When $<S_z>_{av}$ for $Nd^{3+}$ and $Gd^{3+}$ are constrained to remain in phase, the respective $<\mu_z>_{av}$ values make opposing contributions to the bulk magnetic response of an admixed alloy[2].

**II. Samples and Experimental Details**

Polycrystalline samples in the pseudo-binary series $Nd_{1-x}Gd_xRh_3B_2$ (x = 0.075, 0.15, 0.2, 0.225 and 0.25) were prepared by melting together in an arc furnace the appropriate amounts of the two constituent compounds, $NdRh_3B_2$ and $GdRh_3B_2$. Samples of the end members, $NdRh_3B_2$ and $GdRh_3B_2$, were first made by melting together the stoichiometric amounts of the constituent elements in an arc furnace[14]. The pristine Nd and Gd compounds crystallize in the hexagonal $CeCo_3B_2$ structure with $a$ = 5.4423 Å, $c$ = 3.1139 Å and $a$ = 5.4078 Å, $c$ = 3.1177 Å, respectively. X-ray diffraction patterns of the powdered samples of each of the admixed stoichiometries (0.07 ≤ x ≤ 0.25) were confirmed to index to the $CeCo_3B_2$ structure. Fig. 1 shows a comparison of the X-ray diffraction pattern in $Nd_{0.75}Gd_{0.25}Rh_3B_2$ with those in the end members at x = 0 and x = 1. It is apparent that diffraction lines in the admixed stoichiometry remain equally sharp. As an example, an inset panel (a) in Fig. 1 shows the peak due to (200) plane in the three samples on an expanded scale. The inset panel (b) of Fig. 1 shows a plot of the unit cell volume versus the concentration x in $Nd_{1-x}Gd_xRh_3B_2$ series. The values of the unit cell volume were computed from the respective values of the lattice constants, $a$ and $c$, as determined by refining of the diffraction patterns in different samples. As an example, the difference between the calculated and the observed X-ray data in $Nd_{0.75}Gd_{0.25}Rh_3B_2$ is also displayed in the main panel. The stoichiometries of the admixed alloys were also cross-checked by determining ratios of the Nd and Gd compositions using an Element Analyzer JEOL JSX-3222 system, the measured ratios matched (± 1%) with the targeted stoichiometries.

The DC and AC magnetization and the heat capacity data were recorded using Quantum Design (QD) Inc. Superconducting Quantum Interference Device (SQUID) magnetometers (Models MPMS-5 and SQUID-VSM) and a QD Inc. Physical Property Measurement System (PPMS), respectively. Electrical resistance data were recorded using a home made resistivity set up.

**III. Results**

The three panels in Fig. 2 show the temperature (T) dependences of (a) electrical resistivity in nominal zero field, (b) dc magnetization (M) response in 10 kOe and (c) the heat capacity data ($C_p/T$) in nominal zero field as well as in 50 kOe, respectively, in $Nd_{0.75}Gd_{0.25}Rh_3B_2$.

The curves in each of the three panels in Fig. 2 indicate the occurrence of magnetic transition(s). The ρ(T) data in Fig. 2(a) show a knee like feature near T ≈ 22 K. The $C_p/T$ data in zero field in Fig. 2(c) show the onset of enhancement in the specific heat at T ≈ 48 K, followed by another anomalous increase in $C_p/T$ at T ≈ 23 K. The M(T) data in 10 kOe in Fig. 2(b) display a small undulation in magnetization response near 45 K, followed by a prominent peak centered around T ≈ 22 K. We designate the onset temperature of two transitions at 48 K and 23 K in $C_p/T$ versus T data in Fig. 2(c) as $T_{f1}$ and $T_{f2}$. The temperatures $T_{f1}$ and $T_{f2}$ are also identified in Figs. 2(a) and 2(b). We reckon that observation of onset of faster decrease in ρ(T) near 22 K in Fig. 2(a) is typical of the drop in spin disorder resistivity contribution to ρ(T) due to an underlying magnetic ordering. The presence of a peak like feature in M(T) response at 10 kOe (Fig. 2(b)) at the same temperature suggests the magnetic order to be antiferromagnetic like. The two-peak structure in the $C_p/T$ vs. T response in zero field in Fig. 2(c), and the definitive shifts of the two peaks to the lower temperatures on application of a field of 50 kOe seem to imply a non-trivial thermal evolution in magnetic ordering comprising two phase transformations near $T_{f1}$ and $T_{f2}$, across both of which quasi-antiferromagnetism prevails[16]. If the magnetic ordering(s) near 48 K and 23 K had ferromagnetic like character, then the peak(s) would have broadened out on application of a field of 50 kOe, instead of shifting towards the lower temperature, as is known to happen for an antiferromagnetic transition[16].

To comprehend further the two transitions captured in Fig. 2(c), we show in Fig. 3 the temperature dependences of ac (f = 22 Hz, $h_{ac} \approx$ 2.5 Oe) and dc magnetization in nominal zero field (H ~ 1 Oe). The in-phase ac magnetization ($m'_{ac}$) data in Fig. 3(a) is on a semi-log scale, to enable the identification of a tiny peak, a little below $T_{f1}$ of 48 K. The larger peak in $m'_{ac}(T)$ is, however, centered around 22 K, which is just below $T_{f2}$ of $\approx$ 23 K. An inset in Fig. 3(b) shows the M(T) data above 20 K on an expanded vertical scale to enable the viewing of two crossovers of M = 0 axis at $T_{z2}$ and $T_{z1}$. The main panel in Fig. 3(b) shows another crossover of M = 0 axis at $T_{z3}$. The locations of $T_{f1}$ and $T_{f2}$ determined from Fig. 2(c) have also been marked in Fig. 3(b). *Prima facie*, the field cooled dc magnetization in nominal zero field oscillates across the zero value, criss-crossing it thrice at $T_{z1}$, $T_{z2}$ and $T_{z3}$. The larger peak in $m'_{ac}(T)$ in Fig. 3(a) appears to correlate with the peak in M(T) in the main panel of Fig. 3(b). Note that the larger peak in Fig. 3(b) is followed by crossover of dc magnetization to the metastable negative values, thereby, elucidating the notion of a magnetic compensation (at T < $T_{f2}$) expected in the given stoichiometry. A novel feature in Fig. 3(b), however, is the observation of an additional crossover of the M = 0 axis between $T_{f2}$ and $T_{f1}$ in the inset panel of Fig. 3(b).

Figures 4(a) and 4(b) focus attention onto gradual changes in the dc magnetization response as the applied field is enhanced from 50 Oe to 70 kOe. In Fig. 4(a), the M(T) curves in H = 50 Oe, 140 Oe and 300 Oe display an oscillating response, whereas in Fig. 4(b), the oscillatory response in H = 2 kOe evolves into a broad maximum at H ≥ 50 kOe. The crossing of the M = 0 axis below $T_{f1}$ and $T_{f2}$, respectively, is clearly confirmed in the field cooled M(T) run in 50 Oe in Fig. 4(a). Two distinct stages near $T_{f1}$ and $T_{f2}$ continue to remain evident in the field cooled magnetization response in H = 2 kOe in Fig. 4(b). M(T) curves in H=140 Oe, 300 Oe and 2 kOe also reveal the occurrence of a local minimum at T < $T_{f2}$ (marked as $T^*$ in Figs. 4(a) and 4(b)), corresponding to field-induced turnaround in magnetization of the components involved in generating the magnetic compensation phenomenon across $T_{z3}$. In much larger field of 70 kOe, the observation of a shallow peak like response at T ~20 K can be taken to imprint the transition at $T_{f2}$.

To further explore the transitions at $T_{f1}$ and $T_{f2}$, we show in the main panel of Fig. 5, a comparison of the warm up from 6 K of the remanent magnetization ($M_{rem}$) produced and measured in two field values, namely 50 Oe and 20 Oe, respectively. We have also included in the same panel the field cooled M(T) data in H = 50 Oe. $M_{rem}$ at 6 K was obtained by rapidly reducing the field to a chosen field value, after cooling the sample first to 6 K in 50 kOe. In Fig. 5, note first that the $M_{rem}$ curves in H = 50 Oe and 20 Oe differ appreciably only over a limited temperature interval (~ 18 K to 24 K) encompassing the $T_{f2}$ transition at ~ 23 K. On the basis of $M_{rem}$(T) curves recorded at several other field values (all data not shown here), we can surmise that the said differences scale with the field upto H ~ 100 Oe. In nominal zero field, the $M_{rem}$(T) curve across $T_{f2}$ region would smoothly proceed along, and thereafter it would more or less follow the two M(T) curves displayed in Fig. 5. An instructive feature here is that the sharp drop across $T_{f2}$ of the field dependent part of the remanence signal does not affect the (nearly) field independent remanence signal from 6 K upto about 16 K, and thereafter, from a little above $T_{f2}$, upto near $T_{f1}$. The $M_{rem}$(T) curves, above $T_{f2}$, crossover from positive to negative values a little above $T_{z1}$, identified in the field cooled M(T) curve as the temperature below which magnetization values crossover from stable to metastable negative values. To reiterate, *prima facie*, over a large region between $T_{f1}$ and $T_{f2}$, the warm-up $M_{rem}$ and the cool-down M(T) appear to be phase reversed. The inset in Fig. 5 shows that the metastable negative $M_{rem}$ values in H = 50 Oe (at T > $T_{z1}$) return towards the positive side near 43.5 K. Above this temperature, $M_{rem}$ values continue to increase for a little while. However, as T approaches $T_{f1}$, $M_{rem}$ curve joins upto the cool down M(T) run. The transition at $T_{f1}$, therefore, confirms the crossover to the paramagnetic state, where the magnetization is independent of the thermomagnetic history of the sample.

Fig. 6 shows portions of the magnetization hysteresis loops (between ± 600 Oe) at two representative temperatures, viz, 18 K and 28 K, lying below $T_{f2}$ and in between $T_{f1}$ and $T_{f2}$, respectively. The inset panel (a) in Fig. 6 shows the two loops over a larger field interval (± 18 kOe). The inset panel (b) in Fig. 6 shows the temperature dependence of the half width of M-H loop while crossing the M = 0 axis, which may be notionally referred to as effective coercive field $H_c(T)$ relating to the residual remanence.

An evolution in the shape of hysteresis loops in the low field region rapidly happens on going across $T_{f2}$, it results in an enhancement of residual remanence. However, the overall remanence values even at T < $T_{f2}$ remain only a small fraction of the magnetization signal at 50 kOe. M-H curves at fields above ~ 2 kOe at all T (< $T_{f2}$) notionally remain antiferromagnetic like (cf. curves in inset (a) in Fig. 6). The oscillatory character in $H_c$ versus T plot (in inset (b) of Fig. 6) is noteworthy, its two local minima at ~ 21 K and ~ 35 K correlate with the values of $T_{f2}$ and $T_{z1}$ identified in Fig. 5 for the M(T) curve in 50 Oe. The characteristic of a dip in temperature variation of an effective coercive field across a magnetic compensation temperature appears consistent with such data already reported in another non-homogeneous ferrimagnetic system[17].

**IV. Discussion**

The observations in Figs. 2 to 6 need to be rationalized in terms of the organization/reorganization of antiferromagnetically coupled magnetic moments of $Nd^{3+}$ and $Gd^{3+}$ ions and the contribution from conduction electron polarization (CEP) as a function of field and temperature in a multidomain polycrystalline sample. To strengthen the confidence in the observed two transitions and the low net magnetization values for H ≤ 10 kOe at the stoichiometry $Nd_{0.75}Gd_{0.25}Rh_3B_2$, we show in Fig. 7 the magnetization versus temperature plots at H = 10 kOe in $Nd_{1-x}Gd_xRh_3B_2$ series of alloys, as x increases from 0.075 to 0.225. For temperature less than 20 K, the magnetization signal progressively decreases as x enhances. Further, as x increases from 0.15 to 0.225, a magnetic transition appears to set in at a higher temperature of about 45 K (see the arrow mark in Fig. 7). Note now that the magnetization signal between 38 K and 25 K also progressively decreases as x enhances from 0.15 to 0.225. In contrast to the above stated two progressive suppressions, the magnetization signal above about 50 K can be seen to gradually enhance as the Gd concentration increases from 0.075 to 0.225, which implies that the admixed samples are indeed in the paramagnetic regime above 50 K. Below about 40 K, the antiferromagnetic coupling between the magnetic moments of Gd and Nd prevails for $0.15 \leq x \leq 0.225$, and the same trend gets extended down to x = 0.075 for T < 20 K. It is also useful to realize that in Fig. 7, the magnetization curve for x = 0.225 lies considerably below that for x = 0.20. This emphasizes the trend that the competition between the responses from Gd and Nd moments has become intense between x = 0.20 and x = 0.225. The same trend extends further and at x = 0.25, the given admixed rare earth series reaches the zero magnetization limit, and the alloy $Nd_{0.25}Gd_{0.25}Rh_3B_2$ displays the two quasi-antiferromagnetic transitions near 46 K and 23 K, respectively.

In the pristine $NdRh_3B_2$ ($T_c \sim$ 10-15 K), Nd moments in the ferromagnetic state are oriented in the basal plane of the hexagonal lattice[15]. The nearest neighbor distance between the rare earth moments in the basal plane is however much larger than the corresponding distance between them along the c-axis. The structural feature of quasi one-dimensional chains of compressed rare earth ions along c-axis and the associated electronic structure, a priori imply all physical properties, and especially the magnetic response to be anisotropic. From an analysis of the heat capacity data in a single crystalline specimen of $NdRh_3B_2$, it has been surmised[15] that short range correlations amongst the crystal field split ground state doublet of $Nd^{3+}$ ions perhaps persist far beyond its nominal ferromagnetic temperature (~ 10 K), upto about 50 K. It is, therefore, tempting to conjecture that in $Nd_{0.75}Gd_{0.25}Rh_3B_2$ alloy, the Gd ions having much larger spin value have nucleated a precursor magnetic state near $T_{f1} \approx 48$ K. As the transition at $T_{f1}$ commences, concomitantly the thermal stabilization of Nd moments, antiferromagnetically coupled to Gd moments, is triggered. A full fledged three dimensional magnetic order is expected to set in only below the second transition at $T_{f2}$. Nd moments (thermally averaged) are therefore considered to grow rapidly in magnitude on going across $T_{f2}$, and the magnetic response from them is anticipated to compensate the combined signal from Gd moments and the conduction electron polarization. The full planar anisotropy of the Nd/Gd moments in the basal plane is expected to become effective on going across $T_{f2}$. An observation in Fig. 2(c) that the spin disorder resistivity rapidly drops at $T < T_{f2}$ implies that magnetic dynamics of Nd/Gd moments eventually freezes at the second transition. The 1-D fluctuations effects relating to the correlations amongst Gd/Nd spins along c-axis could be important between $T_{f1}$ and $T_{f2}$.

Our choice of the Nd/Gd stoichiometry in the alloy under study has been such that the magnetization signal remains close to near zero value at low fields across both $T_{f1}$ and $T_{f2}$. Of the three competing contributions to the magnetization response in $Nd_{0.75}Gd_{0.25}Rh_3B_2$, those from the Gd spins and the CEP remain in phase, and tend to reach their limiting values (at a given field) quickly on going across the magnetic transition(s). Considering that we witness crossovers from positive to negative magnetization values at $T_{z1}$ ($< T_{f1}$) and $T_{z3}$ ($< T_{f2}$), it is apparent that the contribution from the (antiferromagnetically linked) Nd moments eventually exceeds those from the Gd moments and the CEP, but, the finite coercive field of the alloy, makes the sample assume metastable negative values at low values of applied field. The crossover of the M = 0 line at $T_{z2}$ (see Fig. 3(b) and Fig. 4(a)) at low fields (H ≤ 140 Oe) is presumably affected by the onset of a sharply rising CEP contribution exceeding the very small net negative magnetization (~ 5.5 x $10^{-3}$ $\mu_B$/f.u. in H = 50 Oe) signal prior to $T_{f2}$. No reorientation of the Nd/Gd moments is expected to happen across $T_{z2}$ during the cool-down procedure.

During the field cooled runs, Nd/Gd moments start attempting to reorient towards the field for 300 Oe ≤ H ≤ 2 kOe, while approaching the $T^*$ values lying below $T_{f2}$ (see a schematic drawn in Fig. 4(b)). Note that in H = 50 Oe, the M(T) curve first shows the crossover at ~ 14 K, followed by a turnaround behaviour at ~ 10 K, thereby indicating the importance of metastability/thermal relaxation effects below ~ $T_{f2}$. As described earlier, during the warm up of $M_{rem}$, Gd moments reorient towards the applied field at about 43.5 K while approaching $T_{f1}$, this notion is schematically displayed in the inset panel of Fig. 5.

Magnetization measurements in the admixed (ferromagnetic) rare earth alloys of the kind studied by us had been performed in abundance in 1960s, however, there does not seem to be any report of apparent repeated/multiple magnetic compensations. One probable reason could be the rarity of the low field magnetization measurements at stoichiometries corresponding to *zero magnetization* in that era. Magnetic compensation and field induced spin-orbit reorientation phenomenon for $Sm^{3+}$ ions has been much investigated in recent years[7,8,10-13,18] in Gd doped $SmAl_2$ matrix ($\mu$ = 0.2 $\mu_B$/f.u., $T_c$ ~ 125 K). The field induced spin-orbit reorientation phenomenon in Sm based alloys has to overcome strong magneto-crystalline anisotropy effects at the $Sm^{3+}$ ion[8,12,18]. In the present $Nd_{0.75}Gd_{0.25}Rh_3B_2$ alloy, the reorientation of Nd and Gd magnetic moments on going across the temperature region of $T_{z3}$ has to overcome moderate magneto-crystalline anisotropy effects (in the basal plane) associated with $Nd^{3+}$ ions[5]. The possible differences in the difficulty of reorientations out of the basal plane for Nd and Gd moments become increasingly more important at high fields (> 10 kOe) in a polycrystalline sample. As per a recent study[15], while the easy axis for Nd/Gd moments in the pristine $(Nd/Gd)Rh_3B_2$ compounds lies in the basal plane, the Gd moments can be driven towards saturation limit for H ∥ c at H > 10 kOe, whereas the c-axis can remain a hard direction for Nd moments even for very high fields (H > 400 kOe).

We think that a comprehensive theoretical understanding of the observed behavior of two quasi-antiferromagnetic transitions and an oscillatory magnetic response at low fields in the admixed $Nd_{0.75}Gd_{0.25}Rh_3B_2$ alloy system requires a careful delineation of the interplay between CEP mediated indirect exchange interactions (between Nd/Gd spins) specific to the details of its anisotropic crystal structure and the crystal electric field effects at the $Nd^{3+}$ ions. We may, however, also add a caveat here. It could be argued that at 25% substitution, $Gd^{3+}$ ions randomly distributed amongst the $Nd^{3+}$ ions can be considered to yield an intertwined system comprising two quasi-distinct phases, to be nominally identified as adjoining clusters of Gd ions surrounded by Nd ions, and the isolated Gd ions trapped in the host matrix of Nd ions. The $NdRh_3B_2$ compound is known to have $T_c \sim$ 10-15K, however, in the given stoichiometry under study, the nominal ordering temperature of the composite alloy has enhanced to $T_{f1}$ of ~ 48K, where the magnetic dynamics of the Gd clusters slows down. The magnetic dynamics of a larger fraction of Nd moments forming the backbone of the host matrix slows down only near $T_{f2}$ of about 23 K. The observation that the spin-flip scattering effectively ceases at $T_{f2}$ implies that locking in of all the magnetic moments in the respective orientations awaits the slow down in the spin dynamics of all the Nd ions. Considering the way the higher temperature transition at ~ 48 K in $Nd_{0.75}Gd_{0.25}Rh_3B_2$ has gradually evolved on progressive enhancement of Gd concentration from x = 0.075 to 0.25 (cf. Fig. 7), we think the cluster description is a less likely proposition, we prefer the earlier description in terms of dominance of 1D-fluactuations effects between $T_{f1}$ and $T_{f2}$ and emergence of a 3D-ferromagnet with a planar anisotropy, specific to the given $CeCo_3B_2$ structure, below $T_{f1}$.

To summarize, we have reported a variety of interesting behaviours, which are emanating from (ferromagnetically) exchange coupled rare earth spins in an admixed rare earth alloy, which is near the zero magnetization stoichiometry and possesses anisotropic magnetic characteristics. Zero magnetization alloys with large spin polarization have potential use in spintronics[7], exploration of their material and physics aspects has widespread interest.

**Acknowledgements**

We would like to thank D. D. Buddhikot, S. Mohan, J. Sinha and N. Kulkarni for some of the measurements and P. L. Paulose, S. S. Banerjee, S. M. Yusuf, G. Baskaran and C. Geibel for fruitful discussions.

**FIGURE CAPTIONS**

Fig. 1. (Color online) Comparison of powder X-ray diffraction pattern in $Nd_{0.75}Gd_{0.25}Rh_3B_2$ with those in samples of $NdRh_3B_2$ and $GdRh_3B_2$. The spectra were obtained using PANALYTICAL X'pert PRO Multipurpose X-ray diffractometer (Cu $K_\alpha$ 1.5406 Å). The three patterns in Fig. 1 have been vertically displaced w.r.t. one another to ensure clarity. The inset panel (a) in Fig. 1 focuses attention onto the width and location of (200) line in the three samples. The inset panel (b) shows variation of unit cell volume with x in $Nd_{1-x}Gd_xRh_3B_2$.

Fig. 2. Temperature variation of (a) electrical resistivity ($\rho$) in zero field, (b) dc magnetization (M) in 10 kOe and (c) heat capacity data ($C_p/T$ vs. T) in zero field and 50 kOe, respectively in $Nd_{0.75}Gd_{0.25}Rh_3B_2$. The magnetic ordering temperatures, $T_{f1}$ and $T_{f2}$, identified from $C_p/T$ vs. T curve in zero field in panel (c) have been marked in panels (a) and (b) as well.

Fig. 3. Temperature variation of (a) in-phase ac (f = 22 Hz, $h_{ac}$ = 2.55 Oe) magnetization (m') and (b) dc magnetization (M) in nominal zero field (H ~ 1 Oe) in $Nd_{0.75}Gd_{0.25}Rh_3B_2$. The transition temperatures, $T_{f1}$ and $T_{f2}$, identified from Fig. 2(c) have been marked in both the panels of Fig. 3. The dc magnetization curve crosses the M=0 axis thrice at $T_{z1}$, $T_{z2}$ and $T_{z3}$, respectively. The inset panel in Fig. 3(b) shows the data between $T_{f1}$ and $T_{z2}$ on an expanded scale, where crossovers at $T_{z1}$ and $T_{z2}$ can be more clearly seen.

Fig. 4. Evolution in dc magnetization curves in $Nd_{0.75}Gd_{0.25}Rh_3B_2$ as applied field enhances from (a) 50 Oe to 300 Oe, and (b) 2 kOe to 70 kOe, respectively. The transition temperatures, $T_{f1}$ and $T_{f2}$, identified from Fig. 2(c) have been marked in different panels of Fig. 4. Note, also, the identification of $T^*$ for curves in H = 300 Oe and 2 kOe as the temperature, where a given M(T) curve turns around after displaying a peak just below $T_{f2}$. Fig. 4(b) also schematically shows the relative orientations of Nd/Gd moments across $T^*$ value in 2 kOe.

Fig. 5. The main panel shows comparison of warm-up of remanent magnetization ($M_{rem}$), from 6 K onwards in fields of 50 Oe and 20 Oe, with the field-cooled magnetization (M) in field of 50 Oe. $M_{rem}$ at 6 K at a given field was obtained on reducing the field from 50 kOe to 50 Oe / 20 Oe, respectively, after initially cooling the sample down to 6 K in 50 kOe. The transition temperatures, $T_{f1}$ and $T_{f2}$, and the zero crossover temperature, $T_{z1}$, have been identified in the field-cooled M(T) run. The inset in Fig. 5 displays a portion of the data on an expanded scale. It also shows the turnaround in relative orientations of Nd/Gd moments w.r.t. the external field (H) as the $M_{rem}$ curve measured in H = 50 Oe crosses over from (metastable) negative to (stable) positive values at a temperature of about T ~ 43.5 K.

Fig. 6. The main panel shows the magnetization hysteresis (M-H) loops over the interval ± 600 Oe at 18 K (T < $T_{f2}$) and 28 K ($T_{f2}$ < T < $T_{f1}$). The inset (a) in Fig. 6 shows M-H curves extending upto ± 18 kOe at 18 K and 28 K. The inset (b) in Fig. 6 displays the plot of half width (($H_+$ - $H_-$) / 2) of the M-H loop, while crossing the M = 0 axis, versus temperature. We have also marked in this inset the locations of $T_{z1}$ and $T_{f2}$, identified in Fig. 5.

Fig.7. Magnetization vs. temperature plots at H = 10 kOe in the polycrystalline samples of $Nd_{1-x}Gd_xRh_3B_2$ (x = 0.075, 0.15, 0.20 and 0.225). The arrow marks the notional position of the higher temperature transition with x ≥ 0.15.

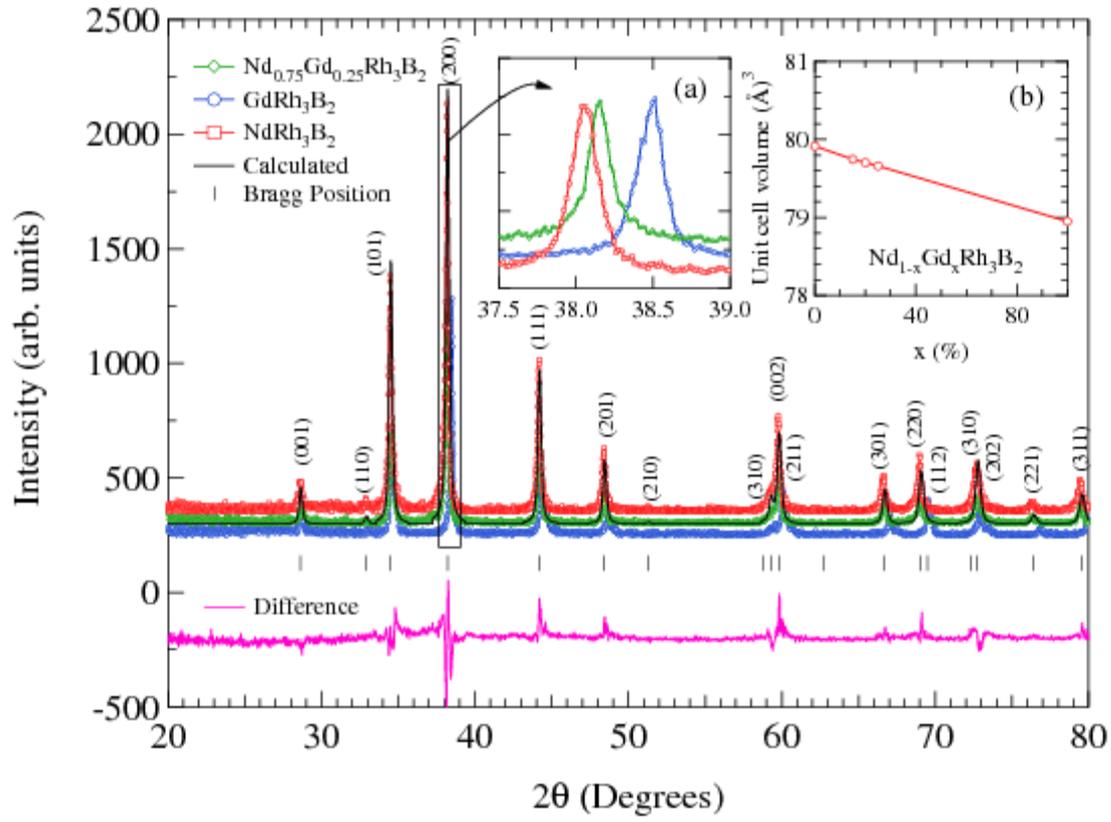

Fig.1.

Prasanna D. Kulkarnt *et al.*

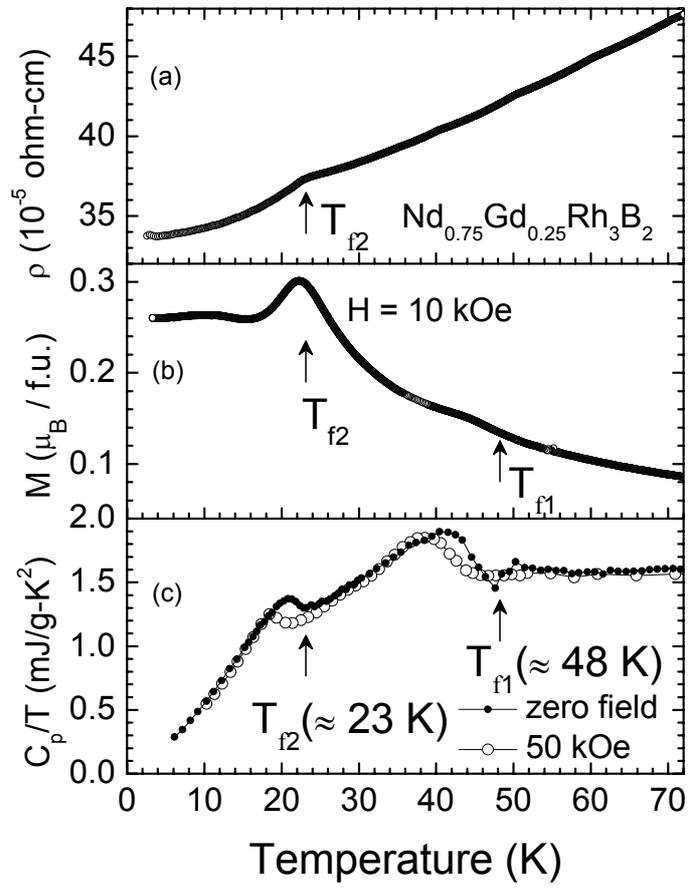

Fig. 2.

Prasanna D. Kulkarni *et al.*

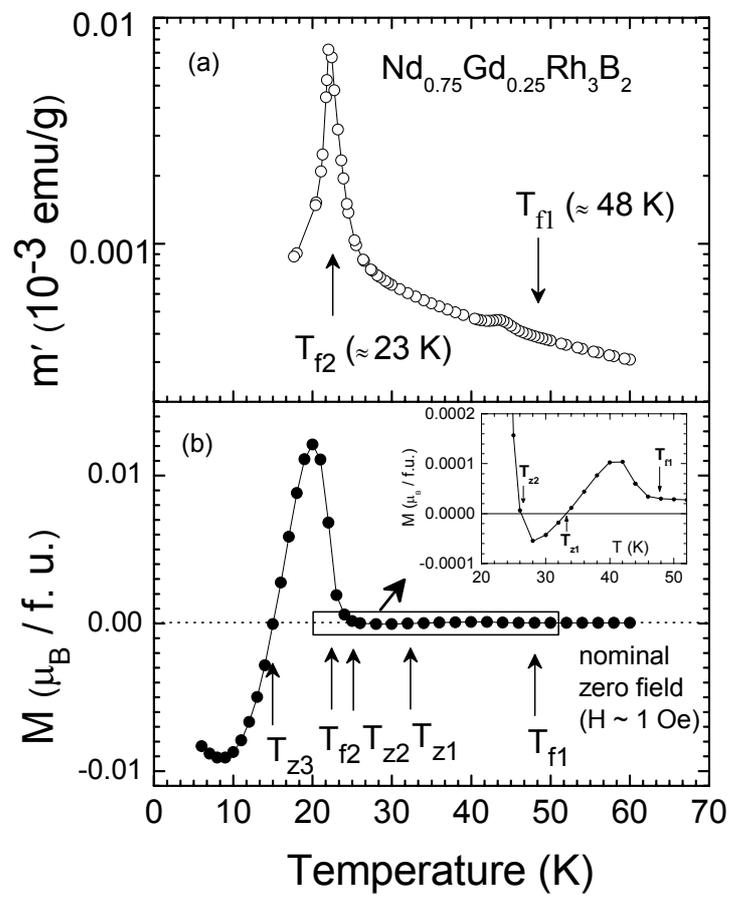

Fig. 3.

Prasanna D. Kulkarni *et al.*

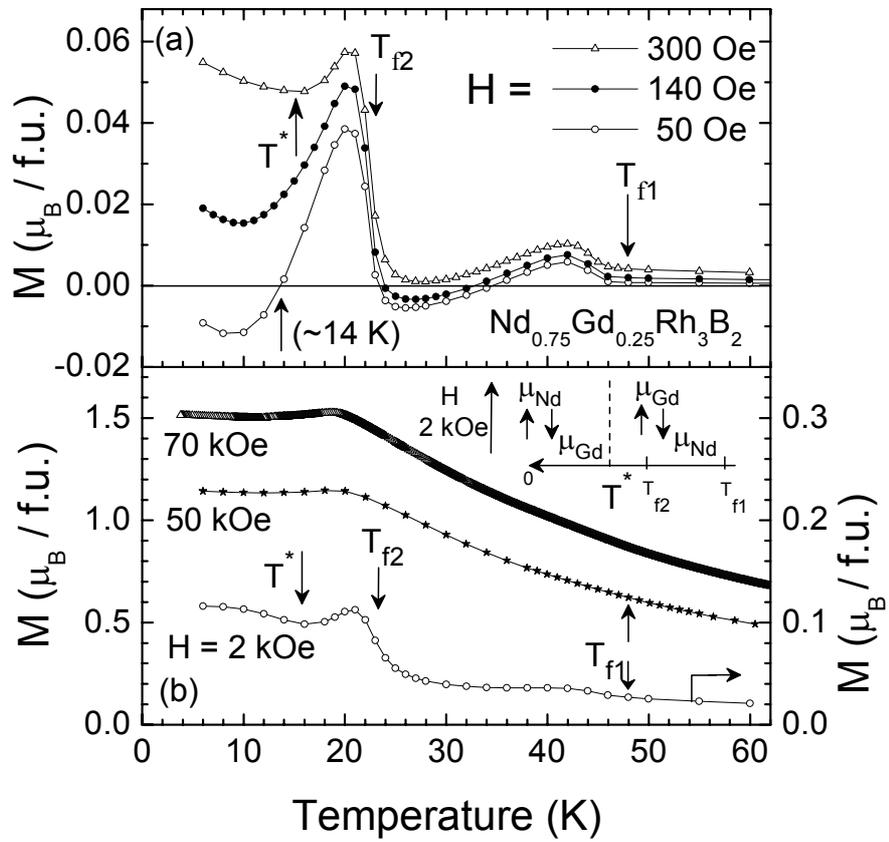

Fig. 4.

Prasanna D. Kulkarni *et al.*

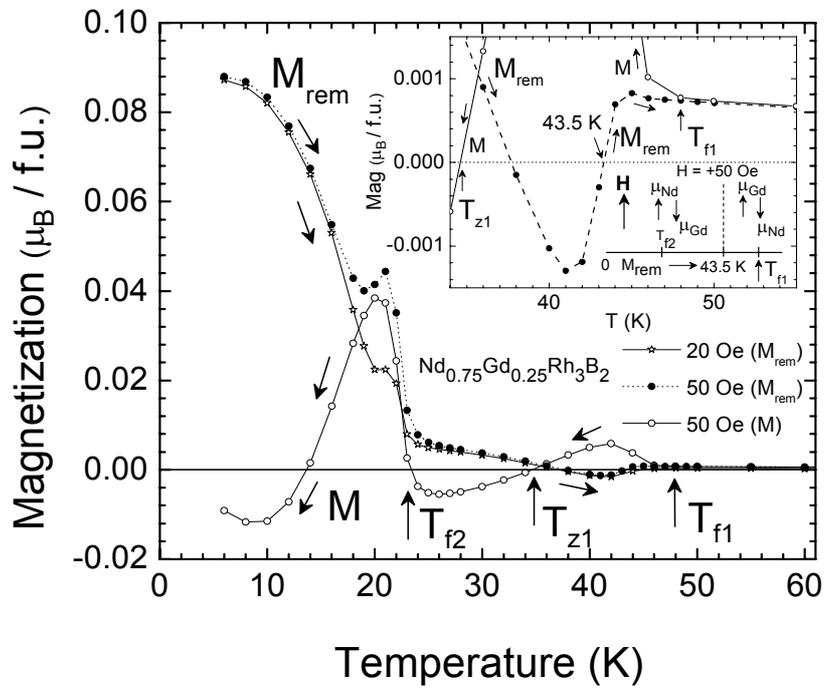

Fig. 5.

Prasanna D. Kulkarni *et al.*

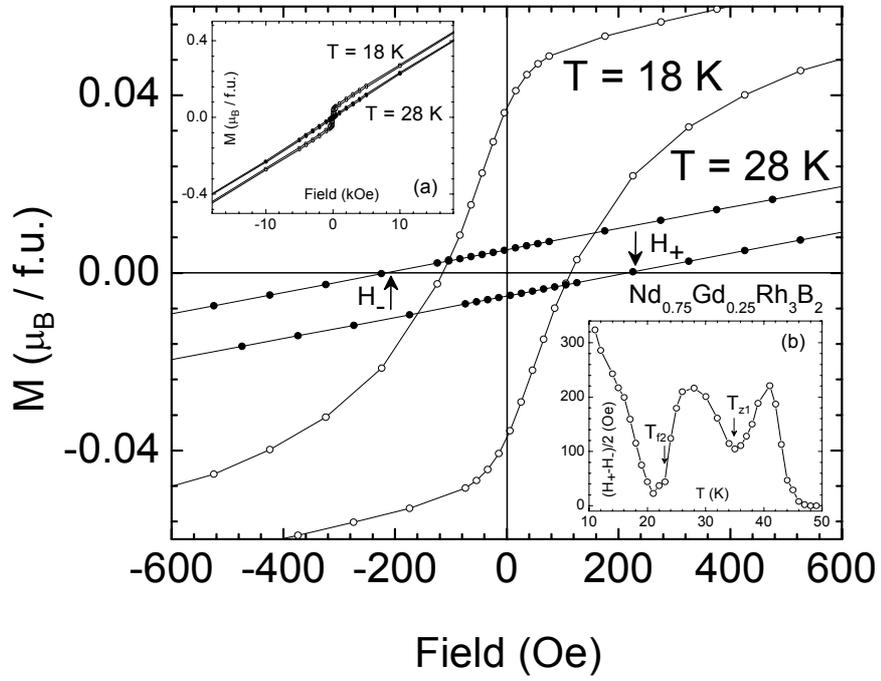

Fig. 6.

Prasanna D. Kulkarni *et al.*

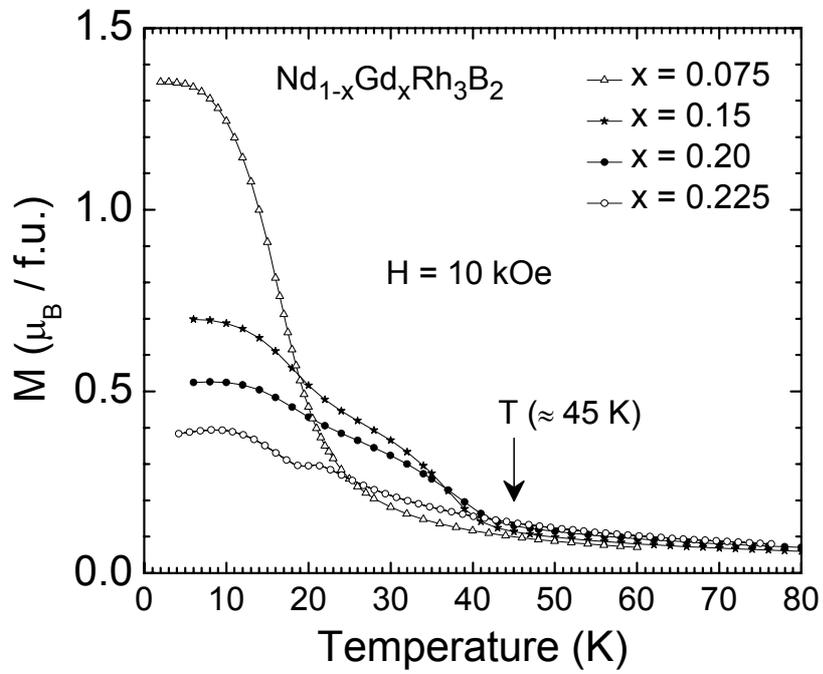

Fig. 7.

Prasanna D. Kulkarni *et al.*